\documentclass[10pt,twocolumn,english,prb]{revtex4-1}
\usepackage[T1]{fontenc}
\usepackage[utf8]{luainputenc}
\usepackage{amssymb}
\usepackage{graphicx}
\usepackage{subscript}

\makeatletter

\@ifundefined{textcolor}{}
{%
 \definecolor{BLACK}{gray}{0}
 \definecolor{WHITE}{gray}{1}
 \definecolor{RED}{rgb}{1,0,0}
 \definecolor{GREEN}{rgb}{0,1,0}
 \definecolor{BLUE}{rgb}{0,0,1}
 \definecolor{CYAN}{cmyk}{1,0,0,0}
 \definecolor{MAGENTA}{cmyk}{0,1,0,0}
 \definecolor{YELLOW}{cmyk}{0,0,1,0}
}

\RequirePackage{colortbl, tabularx}
\@ifundefined{comment}{}
  {
   
  }%

\makeatother

\usepackage{babel}
\usepackage{inputenc}

\makeatother

\usepackage{babel}
\begin{document}

\title{Ultrafast carrier localisation in the pseudogap state of cuprate
superconductors from coherent quench experiments.}

\author{I.Madan$^{1}$, T.Kurosawa$^{2}$, Y.Toda$^{3}$, M.Oda$^{2}$, T.Mertelj$^{1}$,
D.Mihailovic$^{1}$}

\affiliation{$^{1}$Jozef Stefan Institute and International Postgraduate School,
Jamova 39, SI-1000 Ljubljana, Slovenia}

\affiliation{$^{2}$Department of Physics, Hokkaido University, Sapporo 060-0810,
Japan}

\affiliation{$^{3}$Department of Applied Physics, Hokkaido University, Sapporo
060-8628, Japan}
\begin{abstract}
\textbf{A “pseudogap” (PG) was introduced by Mott to describe a state
of matter which has a minimum in the density of states at the Fermi
level, deep enough for states to become localized. It can arise either
from Coulomb repulsion between electrons, or due to an incipient charge
or spin order, or a combination of the two. These states are rapidly
fluctuating in time with random phase, so they are hard to observe
experimentally. Here we present the first coherent quench measurements
of the dynamical transition to the pseudogap state in the prototype
high temperature superconductor }Bi$_{2}$Sr$_{2}$CaCu$_{2}$O$_{8+\delta}$\textbf{,
revealing a marked absence of incipient collective ordering beyond
a few coherence lengths on short timescales at any level of doping.
Instead we find evidence for sub-picosecond carrier localization favouring
a picture of pairing resulting from the competing Coulomb interaction
and strain, enhanced by a Fermi surface instability. }
\end{abstract}
\maketitle
A ``pseudogap'' (PG) was introduced by Mott to describe a state
of matter which has a minimum in the density of states at the Fermi
level, which is deep enough for the states to become localized\cite{MOTT1968}.
It results either from Coulomb repulsion between electrons on the
same atom, a bandgap in a disordered material or a combination of
both. The PG appears in many different systems under current investigation,
and is of ever increasing interest\cite{Timusk1999,Sacepe2010,Mannella2005},
partly due to the resulting unusual physical properties, and partly
because it is commonly associated with the emergence of a long-range
ordered broken symmetry state, such as superconductivity (SC) or a
charge or spin density wave order.\cite{Norman2005} Understanding
its origin in the particular case of high-temperature superconductors
is thought to be of primary importance for determining the mechanism
for the formation of high-temperature superconductivity. In the absence
of evidence for any phase transitions associated with the PG state,
either fluctuating charge-density wave (CDW) or stripe electronic
order has recently been favoured as the origin of the PG state in
the cuprate  superconductors\cite{Comin2014,Parker2010}. The collective
nature of a CDW state would imply critical slowing down associated
with the onset of long range order in the dynamical phase transition,
which would not necessarily be present in the case of stripes. Here
we report on experiments specifically designed to detect critical
behavior associated with the appearance of the PG state in a non-equilibrium
transition with femtosecond coherent control experiments. The method
allows us to either confirm or exclude the existence of long-range
charge or spin order on short timescales and allows us to estimate
the correlation length of the ordered state. Our experiments on a
prototype cuprate Bi$_{2}$Sr$_{2}$CaCu$_{2}$O$_{8+\delta}$ (Bi-2212)
in different regions of the phase diagram show a clear \emph{absence}
of divergence of the single particle relaxation time coincident with
the emergence of pseudogap state, thus excluding the presence of collective
order beyond a few superconducting coherence lengths. The emerging
physical picture for the origin of the pseudogap is Mott-like carrier
localisation and aggregation into short stripes, rather than incipient
dynamic or fluctuating charge density wave order. 

From experiments so far, it is clear that the appearance of the PG
state in the cuprates is not related to a classical thermodynamic
transition, and no signatures of latent heat or anomalies are observed
in the heat capacity temperature dependence.\cite{Loram2000} However,
its gradual appearance below a certain temperature $T^{*}$ is detected
by numerous techniques.\cite{Hufner2008,Norman2005,Timusk1999} The
observed effects are consistent with a depression of the density of
states at the Fermi energy below this temperature. Recently a more
specific origin for the appearance of fluctuating charge density wave
order was discussed \cite{Torchinsky2013} as an alternative to carrier
localization in the form of polarons (or clusters of polarons)\cite{Alexandrov1994,Mertelj2005},
where the origin of the PG in the former case is related to the formation
of a periodic potential opening a gap in the low energy spectrum,
while in the latter, the energy scale associated with the PG is related
to the binding energy of the localized states. The apparent correlation
between the appearance of the PG and stripe order has been intensively
discussed for some time \cite{Bianconi1996,Tranquada1995,Kivelson2003},
and recently reinvigorated \cite{Coslovich2013a,Parker2010,Sugai2006,Torchinsky2013,Comin2014},
opening further questions regarding the relation of stripes with charge
density wave order.

In high-temperature superconducting cuprates, the pseudogap was often
thought to be a precursor rather than a competitor of the SC state.
An apparent 4-fold symmetry of the PG state is compatible with a $d$-wave
SC state symmetry observed by ARPES\cite{Damascelli2003} and STS\cite{Schmidt2011}.
It was however shown that pseudogap does not evolve into the superconducting
gap, but rather coexists with it \cite{Liu2008,Loram2000,Krasnov2000,Coslovich2013,Sacuto2013,Demsar1999,Kabanov1999}.
It was further shown that the PG temperature $T^{*}$ is quite distinct
from the onset of superconducting fluctuations\cite{Madan2014,Rullier-Albenque2006,Li2010,Alloul2010,Kondo2010}.
One similarity which the pseudogap state shares with superconductivity
from the point of view of time-resolved optical experiments is that
in both cases gaps can be destroyed by an ultrashort laser pulse \cite{Toda2011,Kusar2010}.
This fact is especially intriguing, because there is no signatures
of condensation energy associated with the pseudogap observed in the
heat capacity measurements\cite{Loram2000}.

In Bi-2212 the observed charge modulation is incommensurate and has$~\sim4a*4a$
periodicity\cite{Vershinin2004} and it's onset seems to correlate
with the pseudogap temperature $T^{*}$\cite{Comin2014,Parker2010}.
Although pinned in STM measurements, charge order is often thought
to be fluctuating\cite{Torchinsky2013} and consequently should be
characterizable by ultrafast techniques. Whether the pseudogap is
associated with an order parameter characterizing a collective state
with long range order, defined at least on the ultrafast time scale,
or is a stripe-like state of aggregated localized particles is an
important open question (see Fig.1).

\begin{figure}
\includegraphics[width=1\columnwidth]{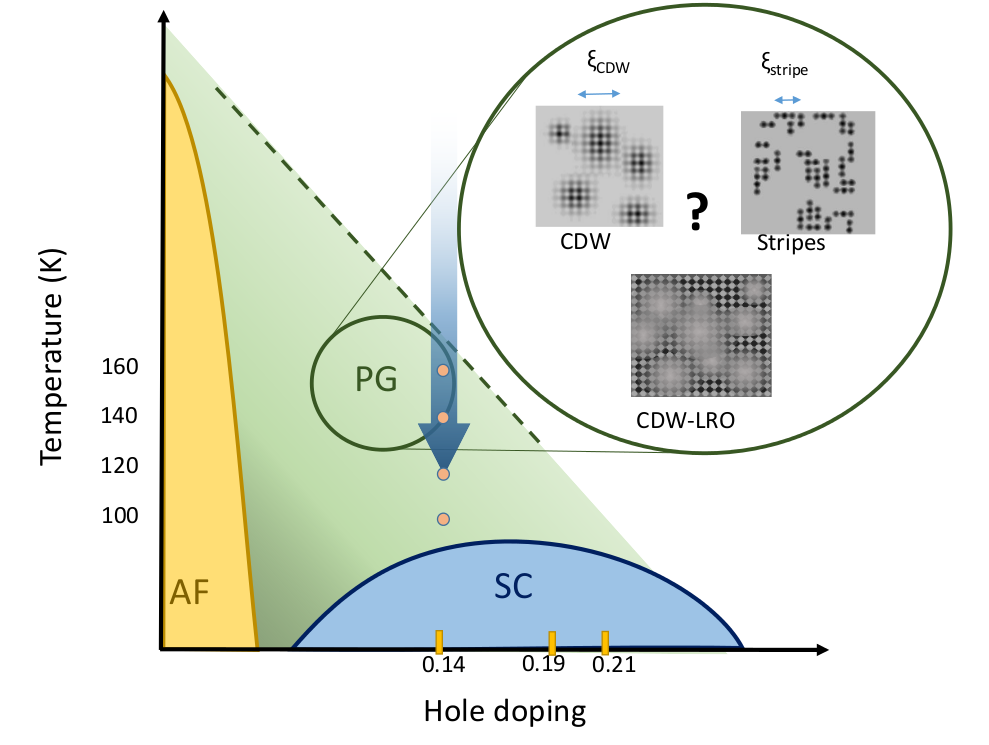}\protect\caption{\textbf{Phase diagram.} A phase diagram highlighting the different
candidates most often discussed for the PG state: a) a CDW-like state
with spatial phase and amplitude fluctuations, b) a stripe-like order
arising from competing interactions, or c) a long-range ordered (LRO)
macroscopically coherent state with spatial amplitude fluctuations.
The arrow shows the quench path. The orange dots schematically represent
the base temperatures at which the data are measured. The doping level
corresponding to the three samples investigated here is also shown.}
\end{figure}

Recent pump-probe experiments measuring the coherent electronic Raman
response have shown broken rotational symmetry associated with the
appearance of the pseudogap state in Bi-2212.\cite{Toda2013} The
pseudogap response has been identified in the $B_{2g}$ symmetry channel
of the parent $D_{4h}$ point group. However these experiments did
not reveal any characteristic length scale or the origin of the PG
excitation - so it was not clear whether the broken symmetry state
is long-range ordered or symmetry is broken only locally by objects
such as polarons. The experiments also did not show whether the symmetry
breaking is static or dynamic.

In gapped laser-excited systems evolving through a transition \textit{in
time}, critical behavior is displayed by the quasiparticle (QP) lifetime
$\tau_{QP}$ which is approximately inversely proportional to the
gap $\Delta$ in the vicinity of the critical time $t_{c}$ of the
transition. This behaviour is ubiquitous and has been seen in both
superconducting and charge-density wave systems appearing as an unmistakeable
divergence of the QP lifetime at $t_{c}$, concurrent with the appearance
of a QP gap. Examples from 3-pulse coherent control experiments include
both quasi-1D and quasi-2D systems: TbTe\textsubscript{3}, DyTe\textsubscript{3},
2H-TaSe$_{2}$, K\textsubscript{0.3}MoO\textsubscript{3}\cite{Yusupov2010}
and superconducting La$_{2-x}$Sr$_{x}$CuO$_{4}$. \cite{Kusar2012}.
The QP lifetime divergences associated with the superconducting transition
are observed ubiquitously in cuprates (YBCO\cite{Kabanov1999}, LSCO\cite{Kusar2005},
BiSCO\cite{Toda2011}, Hg\cite{Demsar2001a}, YBCO124\cite{Dvorsek2002c}
and pnictides \cite{Stojchevska2012}. A similar divergence is also
observed at the SDW transition in pnictides \cite{Pogrebna2014,Stojchevska2010a}.

It is important to emphasize the principal difference of behaviour
of \emph{non-homogeneous systems} in quasi-ergodic slow-cooling experiments
and phase transitions in coherently excited rapid quench experiments
where the elementary excitations may be both highly out of equilibrium
amongst themselves and also spatially inhomogeneous.

Particularly we would like to highlight the difference between the
single particle relaxation time in slow cooling experiments compared
with a system freely evolving in time through a gap-forming transition.
When a system is slowly cooled through the critical temperature, a
gap appears in different regions at different T, and critical behavior
is smeared out and becomes undetectable. On the other hand when a
rapid quench occurs in an inhomogeneous system, a gap starts to appear
simultaneously throughout the whole sample volume and critical behavious
is still observed close to the transition, albeit with a non exponential
relaxation. Thus even if the QP lifetimes are different in different
regions of the sample, a signal revealing critical slowing down will
still be observable. This subtle, but important difference allows
us to search for QP lifetime divergence in the time-resolved optical
reflectivity which would signify the presence of charge order on all
relevant time-scales down to a few tens of fs. The condition for a
coherent response is that the excitation time $\tau_{ex}$ is shorter
than the time it takes for the PG state to form, i.e. the PG recovery
time $\tau_{rec}$. $\tau_{ex}$ is determined by the energy relaxation
time $\tau_{E}\simeq50$ fs \cite{Gadermaier2014}, while here we
will show $\tau_{rec}\simeq600$ fs, so this condition is fulfilled.
The experiments can thus either conclusively confirm or categorically
exclude the presence of long-range order and thus distinguish between
the different PG states shown in Fig. 1 and also enable us to estimate
the effective correlation lengths of PG excitations.

\section*{Results }

\begin{figure}
\includegraphics[width=1\columnwidth]{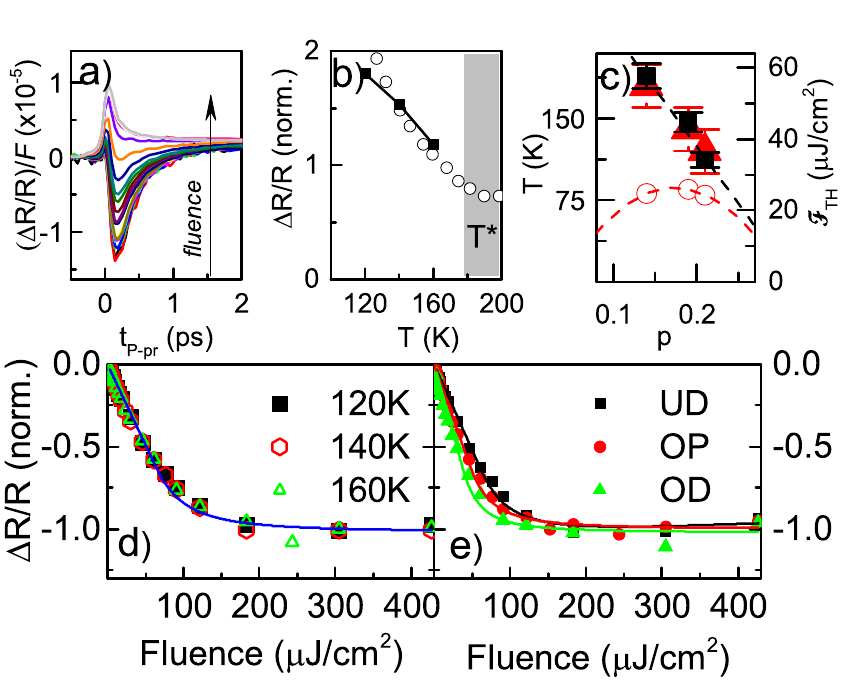}\protect\caption{\textbf{Pseudogap response as a function of fluence.} \label{fig:Fluence}a)
Transient reflectivity normalized by the incident fluence$(\Delta R/R)/\mathcal{F}$
for underdoped sample at 120K. At low excitation the response is linear
and curves overlap. At low excitation negative pseudogap component
is dominant, whereas at high excitation it is saturated and hot-electron
energy relaxation component is prevailing. b) Amplitude of the saturated
pseudogap signal $(\Delta R_{SAT}/R)$ as a function of temperature
(black squares). For comparison renormalized low fluence temperature
dependence of the amplitude from \cite{Toda2011} is plotted (open
circles) c) The value of photodestruction threshold fluence (black
squares) as a function of doping. For comparison doping dependence
of $T^{*}$ (red triangles) and $T_{c}$(red circles) are shown in
the same graph. d) Normalized fluence dependence of the amplitude
of the pseudogap component at different temperatures for the underdoped
sample. e) Normalized fluence dependence of the amplitude of the pseudogap
component for different doping levels: underdoped (UD) at 120 K, optimally
doped (OP) at 120 K and overdoped (OD) at 110 K. }
\end{figure}

The photoinduced reflectivity ($\Delta R/R$) below $T^{*}$ in Bi-2212
ubiquitously shows two relaxation components (Fig.\ref{fig:Fluence}a)).
One is the PG response (which appears as photoinduced decrease in
reflectance) and the other arises from hot electrons energy relaxation
(and appears as an increase in reflectance)\cite{Toda2011,Gay1999}.
With increasing fluence $\mathcal{F}$ both components initially increase
linearly, followed by a saturation of the negative PG component. The
energy relaxation component remains linear with $\mathcal{F}$ for
all measured fluences and does not change significantly with temperature\cite{Toda2011},
so it can be easily subtracted to obtain only the PG signal. The saturation
of the PG response was shown to be associated with the photodestruction
of the PG state\cite{Toda2011} which is clearly nonthermal, since
the estimated lattice temperature rise due to the laser excitation
is < $6$ K.

In Fig. \ref{fig:Fluence}d) and e) we plot normalized fluence dependence
of the amplitude of the PG response for different temperatures and
dopings respectively. For an accurate quantitative analysis of the
fluence dependence and estimation of the threshold fluence (the point
of departure from $\mathcal{F}$-linear behavior) we take into account
the inhomogeneous excitation due to the finite light penetration depth
\cite{Kusar2008}. Fits are represented as solid lines in Fig. \ref{fig:Fluence}d)
and e). 

As shown in Fig. 1d) the fluence behavior nicely scales for different
temperatures indicating that the PG photodestruction threshold fluence
is apparently temperature independent. The value of $\Delta R/R$
at saturation falls with increasing temperature (shown in black squares
in Fig. \ref{fig:Fluence}b)) and follows the T-dependence of the
pseudogap response in weak excitation regime (open circles in Fig.
\ref{fig:Fluence}b)). In Fig. \ref{fig:Fluence}e) we plot the fluence
dependence of the pseudogap response amplitude for the three different
doping levels. The fluence dependence is very similar for all doping
levels, differing only in the threshold value which increases monotonically
with $T^{*}$ and accordingly decreases with doping ( Fig. \ref{fig:Fluence}c)). 

\begin{figure}
\includegraphics{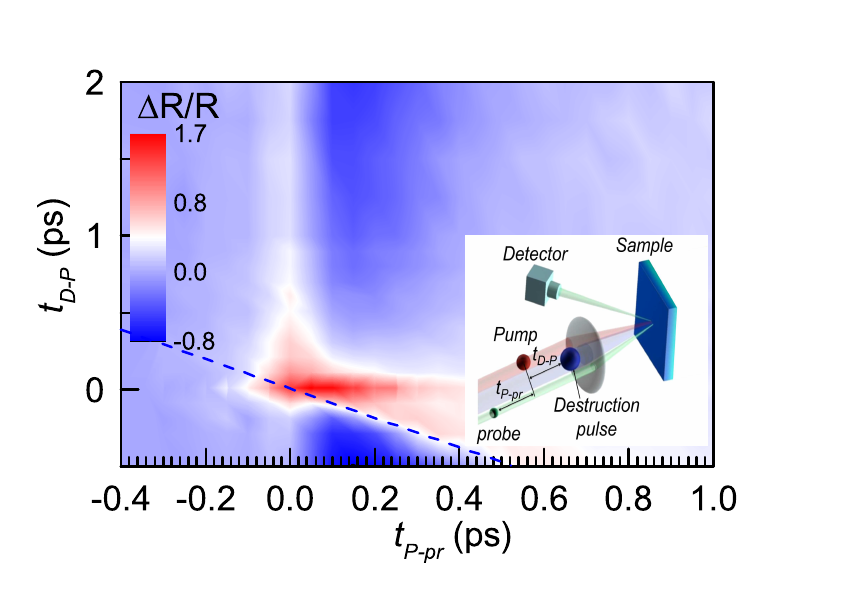}\protect\caption{\textbf{Typical result of the three pulse experiment}. Underdoped
sample at 120 K, the D pulse fluence is 204 $\mu J/cm^{2}$ . The
time of the D pulse arrival is shown by the dashed line. The D pulse
suppresses the negative pseudogap component, whereas the hot electron
energy relaxation response remains intact. Note that at later $t_{D-P}$
the positive component is masked by the stronger negative PG component
with a slower rise time.The inset shows a schematic picture of the
three-pulse pump-probe experiment with the sequence of pulses and
notation of delays. Colors of the pulses do not correspond to their
photon energy.}
\label{Scheme-Colormap}
\end{figure}

\begin{figure}
\includegraphics[width=1\columnwidth]{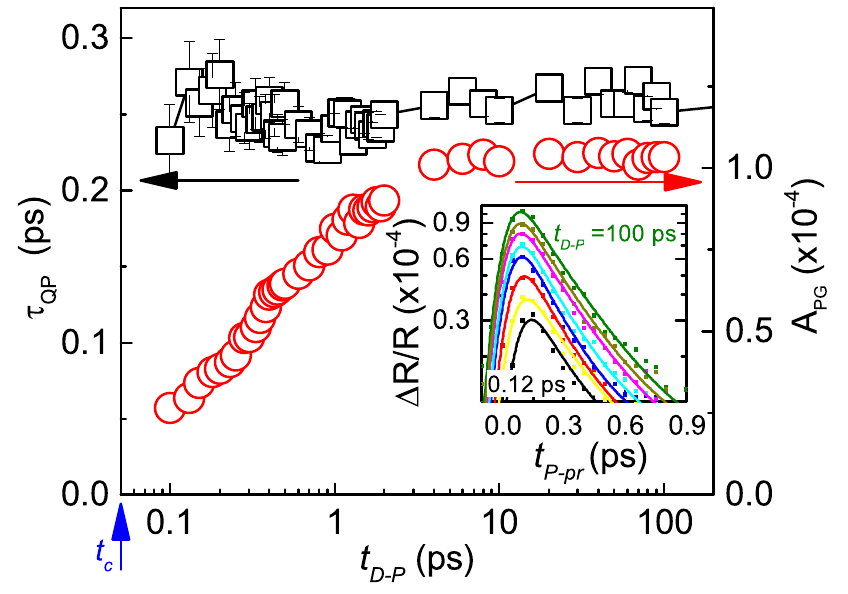}\protect\caption{\label{fig:RelaxationTime}\textbf{Evolution of amplitude and QP relaxation
time.} Quasiparticle relaxation time (black dots, left axis) and the
amplitude(red open circles, right axis) of the pseudogap component
(data shown in inset) as a function of $t_{D-P}$. Quasiparticle relaxation
time remains constant for all values of $t_{D-P}$ . Inset: Pseudogap
component for different values of $t_{D-P}$ ($\mathcal{F}=204\mu\mbox{J/c\ensuremath{m^{2}}},T=120\mbox{K}$).}
\end{figure}

The time evolution of the PG response through the transition measured
in the three pulse experiment is shown in Fig. \ref{Scheme-Colormap}.
First, we note that the \textit{hot electron energy relaxation response}
remains unaffected by the D pulse, (also at room temperature, i.e.
above $T$$^{*}$), and we can thus subtract it in all further data
analysis. The pump induced\textit{ pseudogap response} is suppressed
at the moment when the D pulse arrives (shown by the dashed line).
At later delays $t_{D-P}$ we observe reappearance of the negative
PG response. In the inset of Fig. \ref{fig:RelaxationTime} we plot
the magnitude of the PG signal after the subtraction of the positive
component, measured at the room temperature.  The relaxation times
obtained from the fit to the data in the inset of Fig. \ref{fig:RelaxationTime}
are shown in Fig. \ref{fig:RelaxationTime}. Within experimental error,
the PG relaxation time is constant throughout the PG recovery, at
$\tau_{PG}=0.26\pm0.05$ ps. Remarkably, and in contrast to established
behavior in CDW\cite{Yusupov2010} and superconducting\cite{Kusar2012}
systems investigated so far, no critical behavior of the SP response
is observed close to $t_{c}$, the critical time of the transition.
This observation has important implications with regards to the possible
collective nature of the pseudogap state. We emphasize that the observation
of the absence of critical slowing down at early recovery time is
different from its absence in temperature scans: for possibly inhomogeneous
sample in the case of the temperature dependence at some defined temperature
$T_{1}$ only regions of the sample with this particular $T^{*}=T_{1}$
would show critical slowing down of SP relaxation, whereas in the
quenched case the onset of the pseudogap, and consequently the critical
slowing down, is simultaneous throughout the destruction region. The
absence of divergence might indicate that the normal to pseudogap
transition is of first order. However, in this case we expect to observe
an obvious change of relaxation time at $T^{*}$ \cite{Demsar2002}.
Such effect has not been observed, so a first order transition can
be ruled out.

\begin{figure}
\includegraphics[width=1\columnwidth]{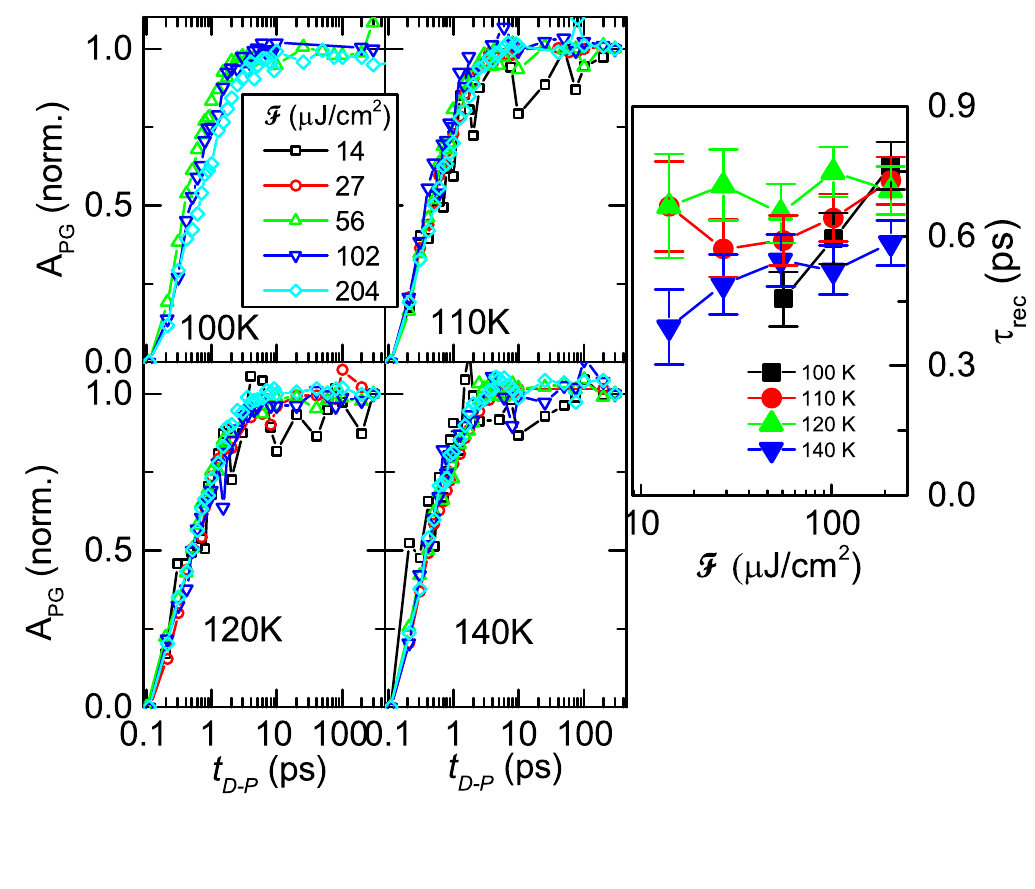}\protect\caption{\textbf{Recovery of the PG state after destruction by a laser pulse
as a function of fluence at different temperatures.} The normalized
amplitude of the pseudogap component as a function of $t_{D-P}$ for
different fluences at a number of temperatures is plotted. Inset:
the recovery time $\tau_{rec}$ as a function of fluence $\mathcal{F}$
at different temperatures. }
\label{Normalized}
\end{figure}

To compare the dynamics of the pseudogap recovery at different destruction
fluences we plot in Fig. \ref{Normalized} normalized amplitude of
the pseudogap component as a function of $t_{D-P}$, at different
temperatures. Within the accuracy of the measurement the PG recovery
time $\tau_{rec}$ is virtually fluence independent. Remarkably, for
all temperatures and fluences, the PG recovery shown in Fig. \ref{Normalized}
can be fit with an exponential function, which is not the case in
the recovery of SC and CDW orders, where the dynamical recovery behavior
associated with the formation of a collective state is more complicated.
Such a simple exponential recovery is consistent with uncorrelated
dynamics of independent particles. 

\section*{Discussion}

The distinct absence of critical behavior as $t\rightarrow t_{c}$
in the PG state gives us new insight into the mechanisms for its formation.
The absence of a divergence in the SP excitation dynamics is a signature
of finite size of the system either limited externally or just indicating
a local nature of the excitation. The $50\:\mbox{fs}$ uncertainty
of the measured value of $\tau_{QP}$ allows us to put an upper limit
on the correlation size of the pseudogap excitation. Taking the Fermi
velocity as a maximum fluctuations propagation speed of $v_{F}\sim150\,\mbox{nm/ps}$\cite{Vishik2010}
we obtain $\xi_{cor}^{max}\sim75$ \AA \, which is only a few superconducting
coherence lengths, and indicates a rather local nature of the PG state.

A number of experiments suggest that the pseudogap is associated with
bound (or localized) states \cite{Zhao1997,Zhao1998,Mihailovic2002,Kohsaka2008,Sakai2013,Coslovich2013a}.
A simple but plausible picture \cite{Kabanov1999} is that photoexcitation
leads to the excitation of carriers from these states into itinerant
states. Thereafter binding takes place on a timescale given by $\tau_{rec}$
which is nearly independent on fluence and temperature, again suggesting
non-collective behavior . This picture is supported by the fact that
the pseudogap is \emph{filled rather than destroyed (i.e. closed)}
after photoexcitation \cite{Smallwood2014}, i.e. a number of delocalized
``in-gap states appear'' without strongly altering the binding energy.
This is tantamount to saying that the states do not act cooperatively,
and there is no change of the energy scale, as the system evolves
through $t_{c}$ in time. 

In the analysis of the fluence dependence {[}Fig. 1d) and e){]} we
have used a model which assumes that the photoinduced absorption is
proportional to the density of photoinduced quasiparticles. This is
in turn linear with excitation intensity. This model has been shown
to give a good description of superconducting condensate destruction,
where the number of particles in the condensate is final \cite{Kusar2008}.
In the case of the pseudogap with localized excitations, the simplest
way to describe the state is in terms of a two-level system (TLS),
in which case the expected $\mathcal{F}$ dependence might be different.
If photoexcitation directly excites particles from the ground state
into the excited state the behaviour as a function of fluence would
be described by saturation of the excited state population $\sim\mathcal{F}/(1+\mathcal{F}/\mathcal{F}_{0})$,
where only half of the localized particles are excited at high intensities.
However, 1.5 eV photons do not excite particles accross the TLS directly,
but create large numbers of electrons and holes through avalanche
multiplication associated with hot carrier energy relaxation which
then populate the excited state and deplete the ground state (filled
with holes). If the hot electron energy relaxation time $\tau_{e-ph}\lesssim50\:\mbox{fs}$\cite{Gadermaier2014}$^{,}$%
\footnote{In our experiment this component is resolution limited and appear
slower. Note that there is another slower component related to thermalization
of hot phonons\cite{Perfetti2007} %
} is shorter than the excited state relaxation time $\tau_{PG}\simeq260$
fs, then a bottleneck is formed, and we revert to the same scenario
for the saturation of the photoexcited response as for the superconducting
condensate. Additional confirmation that localized carriers are not
excited directly by pump photons, but rather as a result of the avalanche
process is that the pseudogap component is slightly delayed with respect
to the hot electron relaxation response - see Fig. \ref{Scheme-Colormap}
(the positive component preceeds the negative component). 

Within this model, the photodestruction fluence threshold $\mathcal{F}_{\mathrm{TH}}$
is proportional to the density of bound carriers and inversely proportional
to the photoexcitation efficiency which is defined as a ratio of energy
spent on quasiparticle excitation to the total absorbed photon energy.
The number of states involved in pseudogap formation can be estimated
from electronic heat capacity and has been shown to decrease with
doping\cite{Loram2000} which explains the proportionality $\mathcal{F_{\mathrm{TH}}\sim\triangle}$
shown in Fig. \ref{fig:Fluence} c). The apparent independence of
the photodestruction fluence on temperature can be explained by relative
inefficiency of thermal excitation from localized states at the temperature
of the measurements, so that number of photoexcited carriers is an
order of magnitude larger than thermally excited. Estimated $\lesssim3\%$
difference in photodestruction energy due to the distinction in the
number of bound carriers at highest and lowest temperatures of the
measurements is within the $\sim6\%$ error of our experiment.

Regarding the origin of the localization, while CDW patches \cite{Torchinsky2013}
and mesoscopic stripes are sometimes confused, and may appear to be
similar in some experimental observations, their implied origin is
different. In the former case, the CDW and accompanying translational
symmetry breaking is caused by the Fermi surface (FS) physics, such
as nesting at specific wavevectors leading to a FS instability. Recent
ARPES studies have revealed that the wavevector of modulation does
not correspond to the nesting between parallel sheets of FS at the
antinodes but rather to the vector connecting ``endpoints'' of ``Fermi
arcs'', so no ``true'' nesting takes place.\cite{Comin2014} Mesoscopic
stripe textures on the other hand are usually considered within the
strong coupling picture to be a result of the competing interactions,
such as microscopic strain caused by localised holes and the Coulomb
repulsion between them, \cite{Kabanov2006,Mertelj2005,Mertelj2007,Keller2008,Alexandrov1994},
or the hole kinetic energy competing with the Coulomb interaction
within Hubbard or $t-J$ models\cite{Agrestini2003,Mishchenko2004,Kivelson1998,He2011}.
However, these models per-se do not address the large variation in
$T_{c}$s observed in the cuprates. Although not directly responsible
for CDW formation, even a weak FS instability may act to additionally
stabilize hole pairs within the strongly correlated picture discussed
above. This adds an additional material-specific component which has
a direct effect on $T_{c}$, namely the degree of nesting of the states
at the Fermi surface\cite{Damascelli2003}. This enhancement is not
necessarily static: short stripe segments may form as dynamically
fluctuating Friedel oscillations around localised carriers (Fig. 6).
Indeed, the $q$-vector observed in X-ray experiments corresponds
closely with the $q$-vector of the lattice and spin anomalies observed
by inelastic neutron scattering experiments in YBa$_{2}$Cu$_{3}$O$_{7-\delta}$
( \cite{Mook1999,Mihailovic2001a,McQueeney1999}). Such a mechanism
would have an effect in enhancing the pair stability in the antinodal
directions (along the Cu-O bonds), and hence raising $T_{c}$.

\begin{figure}
\includegraphics[width=1\columnwidth]{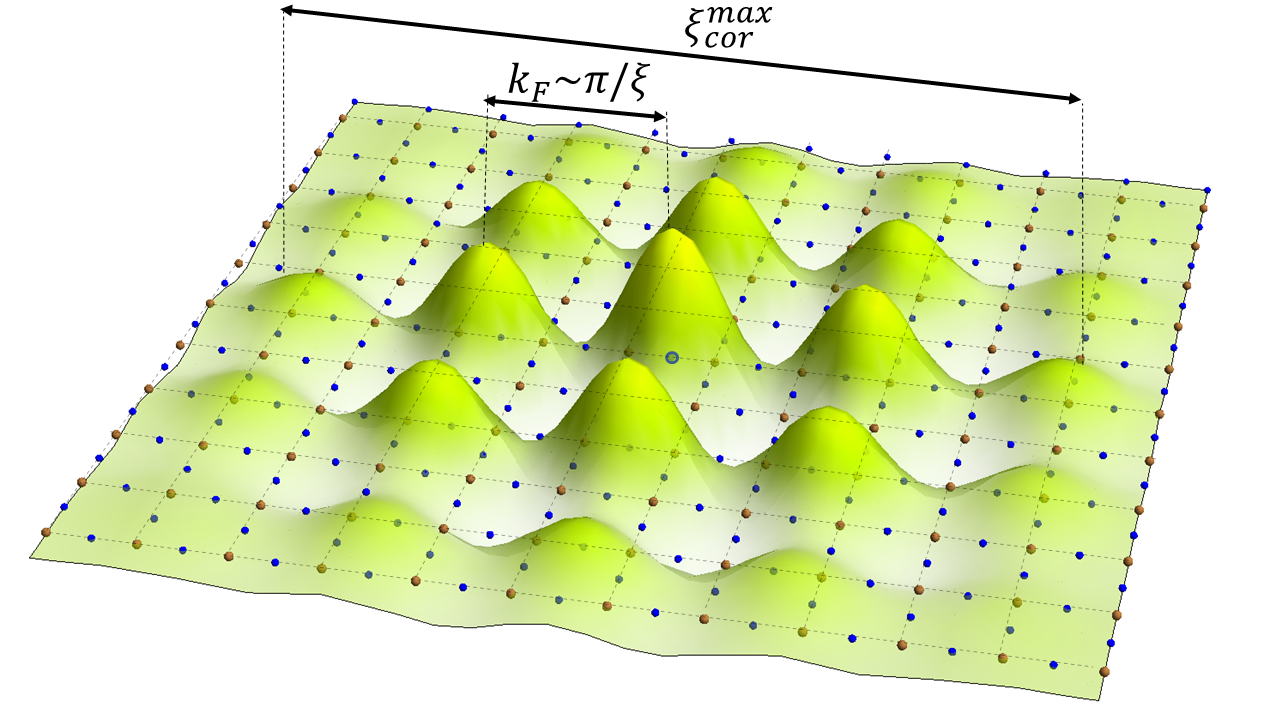}\protect\caption{The real-space charge density map corresponding to a short-range Friedel
oscillation around a localised carrier arising from an enhanced pairing
susceptibility at inter-particle distances of $k_{F}\sim\pi/\xi$.
The correlation length $\xi_{corr.}^{max}$ from the present experiments
is also shown.}

\end{figure}

To conclude, the data presented here, particularly the behaviour of
the PG relaxation time through the coherently excited dynamical transition
in Fig. 4 imply a picture where the pseudogap state is characterized
by a short-range correlated localized carriers, pairs or very small
clusters, locally breaking rotational symmetry\cite{Toda2013}, rather
than proper charge density wave segments (Fig. 1) discussed in the
LaSrCuO system away from 1/8 doping.\cite{Torchinsky2013} The observations
are thus more consistent with a polaronic picture than a dynamically
fluctuating charge density wave with long range order. Our experiments
also cleraly show that the character of the PG state does not change
with doping in Bi2212. Only the energy associated for its destruction
diminishes with increasing doping, reflecting the change of localisation
energy (and $T^{*}$) with increasing screening.

\section*{Methods}

The pulse train of 50 fs 800 nm laser pulses from a Ti:Sapphire regenerative
amplifier with a 250-KHz repetition rate was used to perform pump-probe
(P-pr) reflectivity measurements. For the three pulse experiment each
laser pulse was split in three: the strongest destruction (D) pulse
was used to destroy the state while the evolution of the state was
monitored by measuring the pump-probe response at different delays
between the D and P pulse.

Three samples with different doping levels were investigated in this
work: under- ($T_{c}=81$ K, $T^{*}=180$ K), near optimally- ($T_{c}=85$
K, $T^{*}=140$ K) and over- ($T_{c}=80$ K, $T^{*}=120$ K) doped
Bi2212 with hole concentrations $p=0.14$, 0.19 and 0.21 respectively.
Samples were grown by the traveling solvent floating zone method.
Their critical temperatures were obtained from susceptibility measurements,
doping levels and pseudogap temperatures were estimated from previous
studies\cite{Oda1997}.

\subsection*{Acknowledgments}

$ $

Authors would like to thank Viktor Kabanov for valuable discussions.

\end{document}